\documentclass{article}
\usepackage{amsmath, pb-diagram, lamsarrow, pb-lams, hyperref}

\def\build#1_#2^#3{\mathrel{\mathop{\kern 0 pt#1}\limits_{#2}^{#3}}}

\usepackage{graphicx}
\usepackage{tabularx}
\usepackage{float}

\begin{document}
\title{The framework for simulation of dynamics of mechanical aggregates.}
\author{Petr R. Ivankov, Nikolay P. Ivankov}
\renewcommand{\sectionmark}[1]{}
\maketitle

\begin{abstract}
{A framework for simulation of dynamics of mechanical aggregates has been developed. 
This framework enables us to build model of aggregate from models of its parts.
Framework is a part of universal framework for science and engineering.
}  
\end{abstract}

\section{Introduction}

A set of engineering problems are concerned with aggregates simulation. As a rule models of every part of aggregate are simple. 
However model of whole aggregate may be very complicated. That is the purpose to which the framework has been developed. You can download source code and 
evaluate examples from \url{http://www.genetibase.com/universal-engineering-framework-9.php}. It is worth to note that typical engineering
problems have many aspects that lay outside pure mechanics. Therefore this framework is a part of the universal engineering framework
\url{http://www.genetibase.com/universal-engineering-framework-1.php}, that enables us to simulate complicated engineering phenomena.

\section{Math Background.}

Described framework has a simple background. There exists a set of connected parts of aggregates. If we have two parts connected at the point $A$
then linear $a$ and an angular $\epsilon$ accelerations of one part at point $A$ coincides with corresponding accelerations of the. 
If one part acts on another one with force $F$ and momentum $M$ then the second part acts to first one with force $-F$ and momentum $-M$. 
This is set of conditions is sufficient for construction of aggregate equations. Let us consider that aggregate contains $n$ parts numbered by
$1,..., n$ and $S$ is a set of such pairs $(i, j)$ that $i$-th part is connected to $j$- th one. We shall consider only the case when the graph corresponding to 
$S$ \cite{graphtheory} is a forest \cite{forest}.
According to Lagrangian mechanics \cite{lagrangianmechanics} vector $q$ of Generalized coordinates \cite{generalizedcoordinates} of $i$ -th part 
satifies to the following ordinary differential equation:

\begin{equation}
\ddot{q_i}=A(q_i, \dot{q_i}) + Q_i. 
\end{equation}

Where $A(q_i, \dot{q_i})$ is a part's specific term and $Q_i$ is a generalized force.  We can decompose generalized force in the following way: 

\begin{equation}
Q_i = Q_{i_0} + \sum_{(i, j)\in S}Q_{ij}. 
\end{equation}
Where $Q_{i_0}$ is an external force and $Q_{ij}$ is generalized force caused by action of part $j$ to part $i$. Let us denote vector $W$ as: 
\begin{equation}
W_{ij} = \begin{pmatrix} F_{ij}\\
		M_{ij}\end{pmatrix}.
\end{equation} 
Where $F_{ij}$ and $M_{ij}$ are ordinary force and momentum of action of $j$ - th part to $i$ - th one. Then we have

\begin{equation}
Q_{ij} = B_{ij}(q_i, \dot{q_i}) W_{ij}.
\end{equation} 

Where $B_{ij}$ is a connection specific matrix. Let us denote vector $w_{ij}$ as

\begin{equation}
w_{ij} = \begin{pmatrix} a_{ij}\\
		\epsilon_{ij}\end{pmatrix}.
\end{equation} 

Where $a_{ij}$ and $\epsilon_{ij}$ are linear and angular acceleration at the place of connection between $i$ - th and $j$ - th part.
Then
\begin{equation}
w_{ij} = C_{ij}(q_i, \dot{q_i}) + D_{ij}(q_i, \dot{q_i})\ddot{q_i}. 
\end{equation} 

Using previous equations we can obtain a system of linear equations whose variables are vectors $\ddot{q_i}$ $i\in\{1,...,n\}$ and $W_{ij}$ $(i, j)\in S$.
Resolving of these equations give as $\ddot{q_i}$ and full system of differential equation of aggregate. Note that if we use numerical methods
to solve such equations we should the latter should be normalized. It means that we should make such corrections to generalized coordinates that coordinates and velocities 
 in $i$ - th and $j$ - th part's connection point should be equal for all $(i,j)\in S$. 
 \section{Program Implementation.}
 All equations of previous section are elementary exercises for first year engineering student. The main
 advantage of this work is its program implementation. This advantage may be exhibited by the following example. Let us consider
 a spacecraft with two nonrigid photovoltaics and three flywheels (See Figure 1). 
\begin{figure}[h]
\begin{center}
\hspace{-1cm}
\includegraphics[scale=0.5]{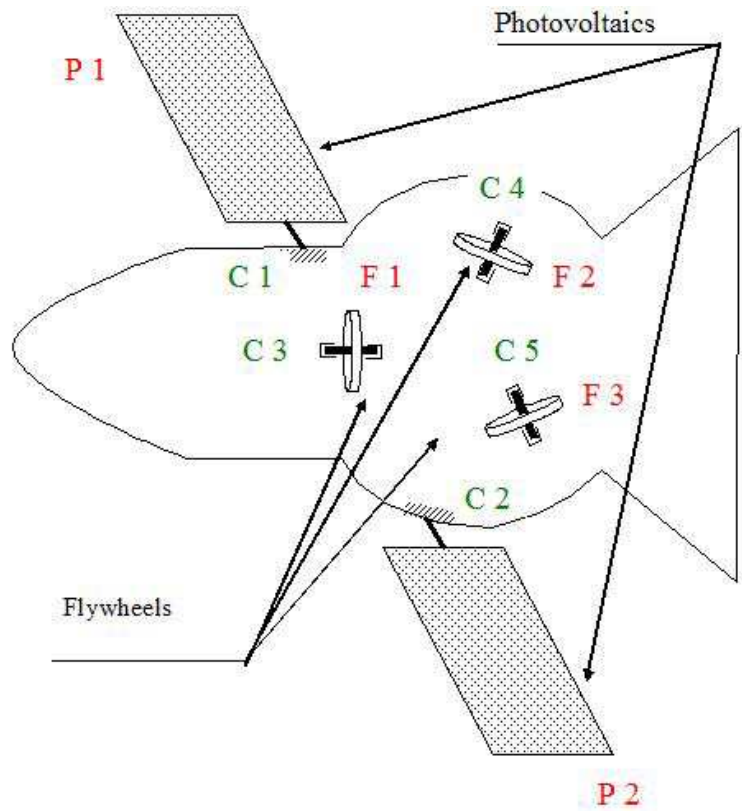}
\caption{A spacecraft with two photovoltaics and three flywheels}
\end{center}
\end{figure}

 This example contains spacecraft with 5 connections \textbf{C 1}, ..., \textbf{C 5}. Flywheels \textbf{F 1}, \textbf{F 2}, \textbf{F 3} and 
 photovoltaics \textbf{P 1}, \textbf{P 2}, \textbf{P 3} are 
 connected to the spacecraft. We can obtain full mechanical model of the spacecraft at once using designer (See Figure 2).
\begin{figure}[h]
\begin{center}
\hspace{-1cm}
\includegraphics[scale=0.5]{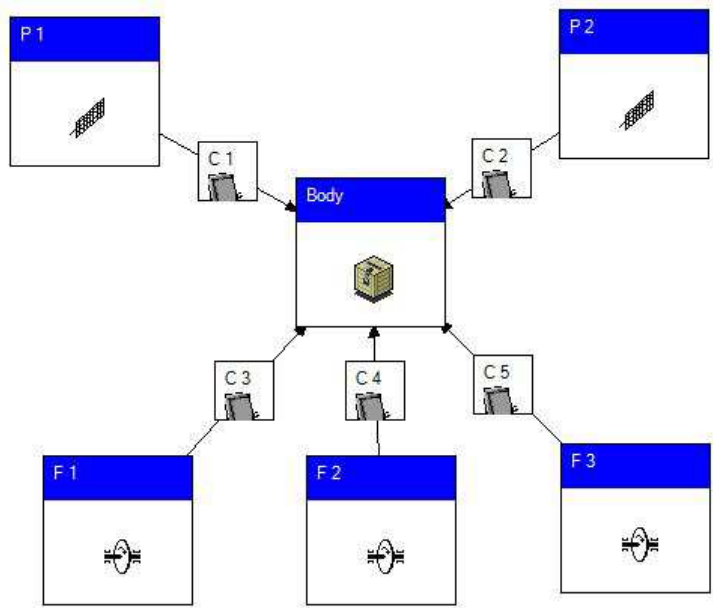}
\caption{A full model of specectaft}
\end{center}
\end{figure}

The idea of software is very simple. All mechanical objects should implement interface \cite{interfaces} \textit{IAggregableMechanicalObject}.
You can download source code of this interface from \url{http://www.genetibase.com/universal-engineering-framework-9.php}.
Then designer enables us to construct full models of aggregates from objects those implement this interface.
 \section{Implemented mechanical objects.}
Now three types of mechanical objects are implemented. You can develop your own type of object. To do this you should implement
\textit{IAggregableMechanicalObject} interface. Let us consider types of implemented objects.

\subsection{Rigid body.}
Rigid body is a simplest object of 6D dynamics. It is characterized by mass and moment of inertia.
Key properties of our rigid body implementation are places of connections. 
Connections are defined by those coordinates and orientation. 
You can define those coordinates $X$, $Y$, $Z$ and 
components $Q_0$, $Q_1$, $Q_2$, $Q_3$ of quaternions of orientation.
\subsection{Flywheel.}
Flywheel has general properties of rigid body and additional ones. Additional properties of flywheel include
a moment of inertia of a wheel, initial angular velocity of a wheel and a moment that acts to wheel. 
\subsection{Elastic vibrations body.}
Elastic vibrations body is a mechanical system of infinite degree of freedom. 
Usually math model of this object contains finite degree of freedom with finite set of valuable harmonic oscillations. 
Every harmonic oscillation may be described by following second order ordinary differential equation:
\begin{equation}
A\ddot{q}+\epsilon\dot{q}+cq= Q. 
\end{equation} 
where $q$ is a generalized coordinate and $Q$ is a corresponding generalized force.

\section{Advanced example. Controlled spaceraft.}
Let us consider the following example. We have a spacecraft (See Figure 3).

\begin{figure}[h]
\begin{center}
\hspace{-1cm}
\includegraphics[scale=0.5]{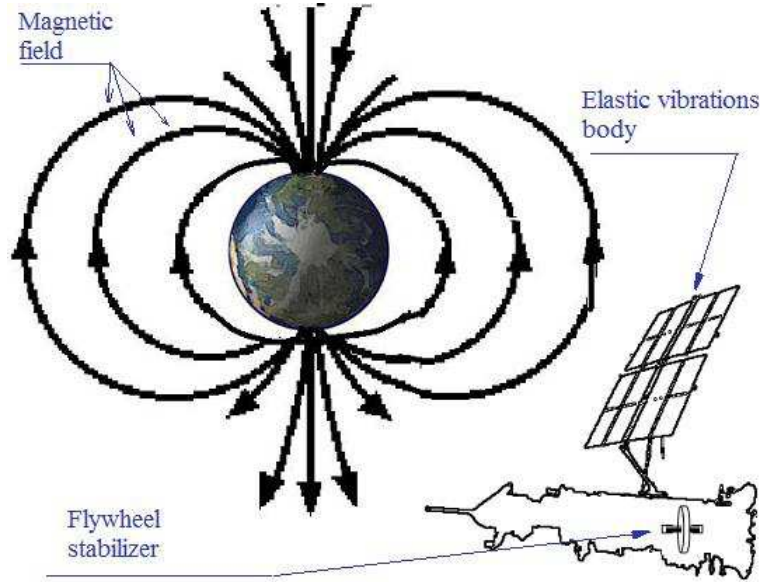}
\caption{A controlled spacecraft.}
\end{center}
\end{figure}

Currents of its equipment interact with Earth's magnetic field. Spacecraft has a photovoltaic that is an elastic vibrations body. 
The is a flywheel that is used for angular stabilization of the spacecraft.
Using the designer we may simulate this situation in the way represented at Figure 4.
\begin{figure}[h]
\begin{center}
\hspace{-1cm}
\includegraphics[scale=0.5]{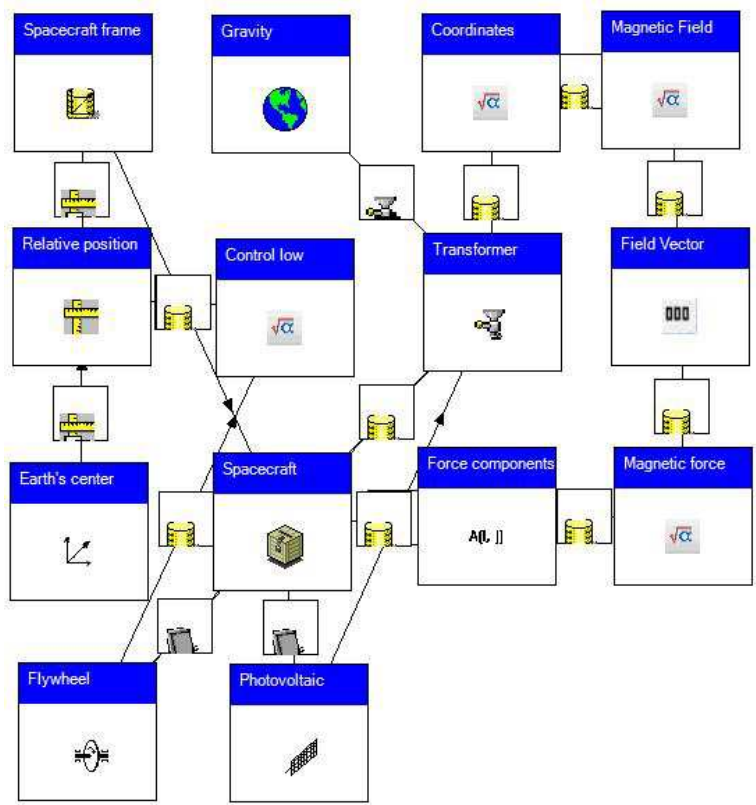}
\caption{A controlled spacecraft design.}
\end{center}
\end{figure}
Let us briefly explain this situation. First of all we setup \textbf{Coordinates of Spacecraft}. 
We need them for definition of gravitational acceleration and magnetic field. Then we construct \textbf{Magnetic Field} 
by formulas. 
It is convenient to represent magnetic field with a vector. Therefore we've used \textbf{Field Vector}.
Its usage enables us to define mechanical moment as a vector product of magnetic induction and 
magnetic moment of spacecraft:
\begin{equation}
M=d\times B. 
\end{equation}
where $d$ is magnetic dipole momentum of spacecraft and $B$ is vector of magnetic induction of th Earth.
Vector product operation is supported by formula editor embedded into framework. The \textbf{Gravity} component is used for 
simulation of gravitational accelerations. Using above components we have defined inertial accelerations and mechanical moments.
Now let us constuct control system. Its first element is a sensor. We will use a sensor of local vertical. 
To do this we define \textbf{Spacecraft frame} and \textbf{Earth's center frame}. 
\textbf{Relative position} (6D) enables us to simulate the sensor of local vertical. 
The \textbf{Control law} uses it. And at last moment of \textbf{Flywheel} is calculated by the \textbf{Control law}.
You can download this example from
\url{http://www.genetibase.com/universal-engineering-framework-9.php} and evaluate it.

\section{Conclusion.}

This framewok is very useful for simulation of complicated phenomena that include mechanical ones.
Authors of this article think that present-day engineering software should (potentialy) include 
all kinds of physical phenomenons. Authors would like to aknowldge the Genetibase company that
kindly sited source code and examples on their site.

\end{document}